\documentclass[
 reprint,
superscriptaddress,
amsmath,amssymb,
aps,
longbibliography
]{revtex4-1}

\usepackage{graphicx}
\usepackage[dvipsnames]{xcolor}
\usepackage{dcolumn}
\usepackage{bm}

\usepackage{hyperref}
\hypersetup{
colorlinks   = true, 
urlcolor     = blue, 
linkcolor    = blue, 
citecolor   = blue 
}

\begin{document}

\newcommand{\TUDelft}{Chemical Engineering Department, Delft University of Technology, Van der Maasweg 9, NL-2629 HZ Delft, The Netherlands}
\newcommand{\UtrecthUniHenk}{Institute for Theoretical Physics and Center for Extreme Matter and Emergent Phenomena, Utrecht University, Princentonplein 5, 3584 CC Utrecht, The Netherlands}
\newcommand{\UtrecthUniDaniel}{Debye Institute, Condensed Matter and Interfaces, Utrecht University, PO Box 80.000, 3508 TA Utrecht, The Netherlands}

\title{Observation of the quantized motion of excitons in CdSe nanoplatelets}

\author{Michele Failla}
\affiliation{\TUDelft}

\author{Francisco García~Fl\'{o}rez}
\affiliation{\UtrecthUniHenk}

\author{Bastiaan B.~V.~Salzmann}
\affiliation{\UtrecthUniDaniel}

\author{Daniel Vanmaekelbergh}
\affiliation{\UtrecthUniDaniel}

\author{Henk T.~C.~Stoof}
\affiliation{\UtrecthUniHenk}

\author{Laurens D.~A.~Siebbeles}
\affiliation{\TUDelft}

\begin{abstract}
We show that the finite lateral sizes of ultrathin CdSe nanoplatelets strongly affect both their photoluminescence and optical absorption spectra.
This is in contrast to the situation in quantum wells, in which the large lateral sizes  may be assumed to be infinite.
The lateral sizes of the nanoplatelets are varied over a range of a few to tens of nanometers.
For these sizes excitons experience in-plane quantum confinement, and their center-of-mass motion becomes quantized.
Our direct experimental observation of the discretization of the exciton center-of-mass states can be well understood on the basis of the simple particle-in-a-box model.
\end{abstract}

\maketitle

\begin{figure}[!h]
\includegraphics[width=0.5\textwidth]{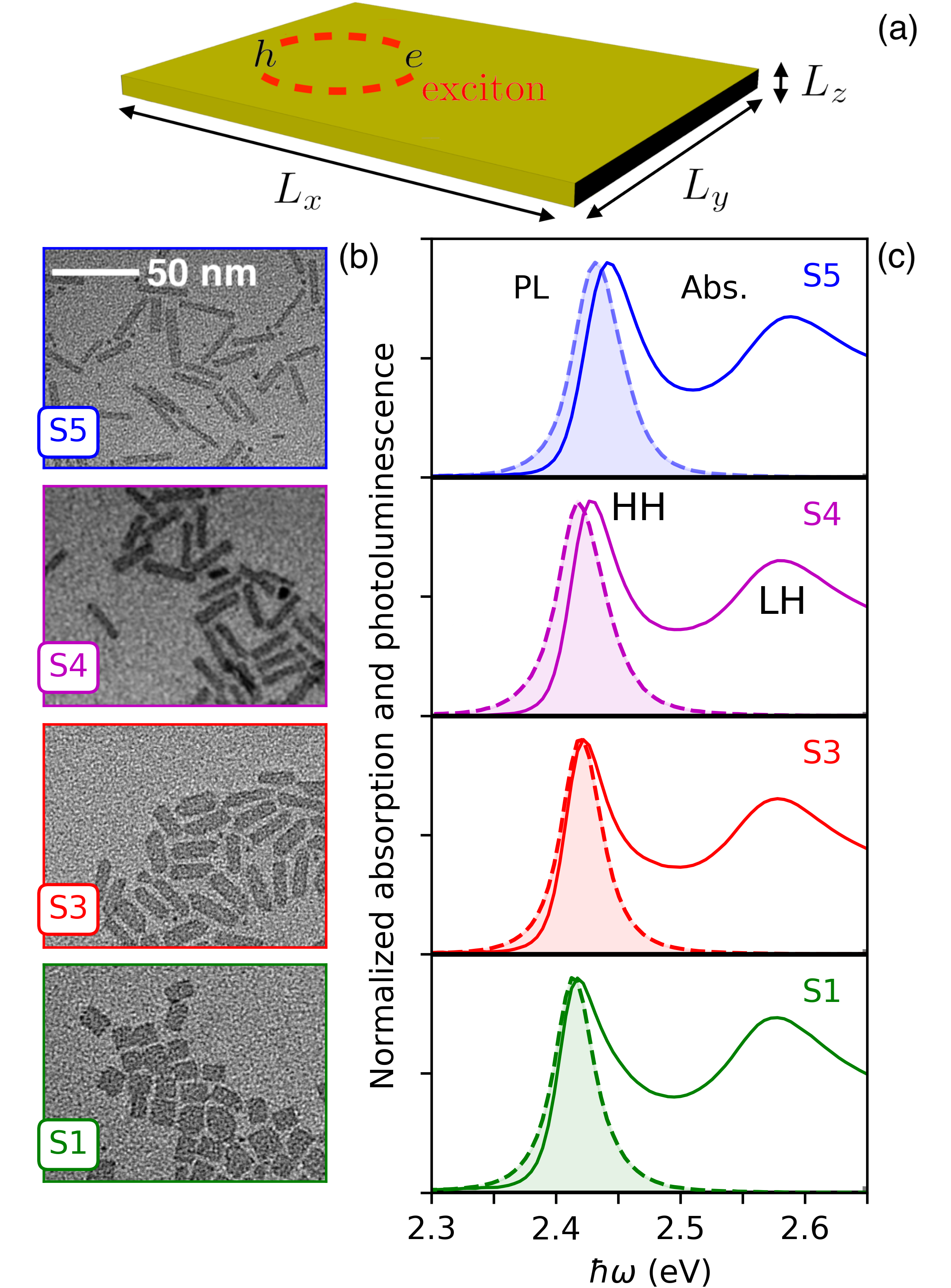}
\caption{(a) Scheme of a nanoplatelet. The exciton center-of-mass motion is influenced by the lateral sizes $L_x$ and $L_y$.
(b) TEM images of the investigated CdSe-NPLs. 
(c) Photoluminescence (dashed lines) and optical  absorption (solid) spectra for samples in (b).  
The absorption peaks at low and high energies are due to HH and LH excitons, respectively.  
}
\label{fig1}
\end{figure}

Cadmium selenide nanoplatelets (CdSe-NPLs) are solution-processable two-dimensional semiconductor nanomaterials which are grown with an atomically precise thickness \cite{Ithurria:2008,Yu:2020}, see Fig.\,\ref{fig1}(a) and (b). 
They receive considerable attention due to promising optical properties for optoelectronic applications, including strong optical oscillator strength  \cite{Naeem:2015}, large exciton binding energy \cite{Naeem:2015,Achtstein:2012,Tomar:2019}, weak exciton-phonon coupling \cite{Achtstein:2012}, high photoluminescence (PL) quantum yield  \cite{Achtstein:2018}, efficient two-photon absorption \cite{Li:2015,Scott:2015,Heckmann:2017}, and lasing \cite{Grim:2014,Li:2015,Yang:2017}.

The steady-state optical absorption spectrum exhibits well-resolved peaks due to the formation of heavy-hole (HH) and light-hole (LH) excitons, see Fig.\,\ref{fig1}(c).
The energy at which the HH and LH peaks appear in the optical absorption spectrum, as well as the PL energy, strongly depend on the NPL thickness, $L_z$ in Fig.~\ref{fig1}(a), which is of the order of a nanometer \cite{Ithurria:2008,IthurriaNM:2011}.
Hence, in a NPL excitons are strongly quantum confined along this (vertical) direction. 

Besides the atomically precise thickness, the lateral sizes of CdSe-NPLs, $L_x$ and $L_y$ in Fig.\,1(a), can also be tuned  from a few to tens of nanometers by varying the synthesis procedure \cite{Bertrand:2016}.
The strong vertical and  weak, intermediate or strong in-plane   quantum confinement regimes \cite{Rajadell:2017,Richter:2017} make the optical properties  of  NPLs to be in between those of quantum wells (with lateral sizes exceeding microns or larger) and quantum dots. 
Experimentally, reducing the lateral sizes of CdSe-NPLs has been reported  to blue shift and narrow the optical absorption and PL peaks \cite{IthurriaJACS:2011,IthurriaNM:2011,Bertrand:2016}, decrease the absorption cross section \cite{Yeltik:2015} and affect the optical gain and amplified spontaneous emission \cite{She:2015,Yeltik:2015,OlutasACS:2015,LiNL:2017,LiChemi:2018}.

An interesting peculiarity, commonly observed in the absorption spectrum of CdSe-NPLs,  is the asymmetric shape of the HH exciton peak which exhibits a tail at the  side of higher photon energy, see Fig.\,1(c), whose  explanation has been inconclusive.
It has for instance been ascribed to fluctuations in the confining potential of NPLs that lead to localization and consequently an increased energy of the excitons \cite{Schnabel:1992,Grim:2014,Naeem:2015,Tomar:2019}, or to internally excited exciton states \cite{Achtstein:2016, Morgan:2018}.
As depicted in  Fig.\,\ref{fig1}(a),  excitons have a Bohr radius much smaller than the lateral dimensions of a NPL. 
Hence, the centre-of-mass (COM) translational motion in the plane of a NPL must also be considered, and photoexcitation to COM motional states with nonzero in-plane momentum can indeed give rise to the high-energy tail of the HH exciton peak.
Until now, however, this COM motion of an exciton has not yet been invoked to explain the PL and absorption spectra.
The finite lateral size of NPLs then also affects the COM energy of excitons, as discussed by Richter  \cite{Richter:2017}.
The presence of these states has also been invoked to explain the exciton dynamics  probed by transient resonant four-wave mixing \cite{Naeem:2015} and  transient PL spectra   at temperatures below 200\,K \cite{Scott:2019,Specht:2019}.
The latter exhibits two PL peaks with a lateral size-dependent energy difference of tens of millielectronvolts.
Nonetheless, these PL peaks have also been explained by considering phonon-replicas \cite{Tessier:2013}.\\

In this Letter, we directly observe the discretized center-of-mass states of excitons from the lateral size dependence of experimental PL and optical absorption spectra of CdSe-NPLs. 
Both spectra are simultaneously interpreted and accurately modeled by considering the in-plane COM motion of excitons. 
In particular, quantum effects of spatial lateral confinement explain the size dependence of both the PL and absorption peak positions and widths. 
Our work provides an explanation of the hitherto open problem of understanding the effects of lateral size of NPLs on the properties of excitons. It can now be understood how variation of the lateral size of (quasi) two-dimensional materials provides a tool to continuously tune their optical properties, in addition to the discrete changes by adjustment of the thickness.\\

\begin{table}[t]
\begin{ruledtabular}
\begin{tabular}{ccccc}

		Sample &
		$L_x^{\textnormal{avg}}$	&		
		$L_y^{\textnormal{avg}}$	&		
		Area  		&
		$r_l$ 		\\

		  &
		(nm)	&		
		(nm)	&		
		(nm$^2$)	&		
				\\
		
\colrule

		S5
		&$29.3\pm3.3	$				
		&$5.4\pm0.8	$				
		&159
		&5.4 \\
		
 		S4
		&$26.1\pm3.3	$				
		&$6.4\pm1.1	$				
		&167
		&4.1\\ 

 		S3
		&$25.4\pm2.9	$				
		&$8.1\pm0.9	$				
		&204	
		&3.2	\\
					
 		S1
		&$13.7\pm2.2	$				
		&$13.4\pm1.9	$				
		&183	
		&1.0	
		
\end{tabular}

\end{ruledtabular}
\caption{Average lateral sizes ($L_x^{\textnormal{avg}},L_y^{\textnormal{avg}}$), area and aspect ratio, $r_l$, of the  CdSe-NPLs samples.
Samples are labeled based on their aspect ratio.}
\label{tab}
\end{table}

Four samples of CdSe-NPLs with a thickness of 4.5 monolayers   ($L_z\simeq 1.4$\,nm \cite{Achtstein:2012}) and different average lateral sizes were  synthesized by following Ref.\,\cite{Bertrand:2016}, see Supplemental Material Sec.\,I.
The average lateral sizes ($L_x^{\textnormal{avg}}$, $L_y^{\textnormal{avg}}$) were determined from Transmission Electron Microscopy (TEM) images, as shown in Fig.\,\ref{fig1}(b), by assuming a Gaussian size distribution, see Fig.\,SM1 in the Supplemental Material.
The results are shown in Table\,\ref{tab}, together with the NPL areas and aspect ratios $r_l\equiv L_x^{\textnormal{avg}}/L_y^{\textnormal{avg}}$.
The samples are labeled based on their aspect ratio (first column in Table I).
The data presented in this work are obtained from PL and optical absorption ensemble measurements on NPLs dispersed in hexane. 

Fig.\,\ref{fig1}(c) shows  the dependence of PL (dashed lines) and absorption spectra (solid) on the NPL size.
Spectra are normalized for comparison. 
Two  clearly distinguishable  absorption peaks are related to  HH and LH exciton states. 
In agreement with Refs.\,\cite{IthurriaJACS:2011,Bertrand:2016}, by decreasing the aspect ratio, we observe: (i) a red shift  of  all features, (ii) decrease  of the PL linewidth,   (iii) decrease of the Stokes shift, and (iv) increase of the asymmetry (tail at the high-energy side) of the HH exciton peak. 
As mentioned above,  the (iv)-th observation may be related to the presence of phonon replicas  \cite{Tessier:2013}. 
However, the latter cannot be dominant, since the small Stokes shifts in Fig.\,\ref{fig1}(c) imply weak exciton-phonon coupling.
In addition, phonon replicas also give rise to a strong asymmetry at the low energy side of the PL peak, in contrast to the data in Fig.\,\ref{fig1}(c). 

\begin{figure}[b]
\center
\includegraphics[width=0.5\textwidth]{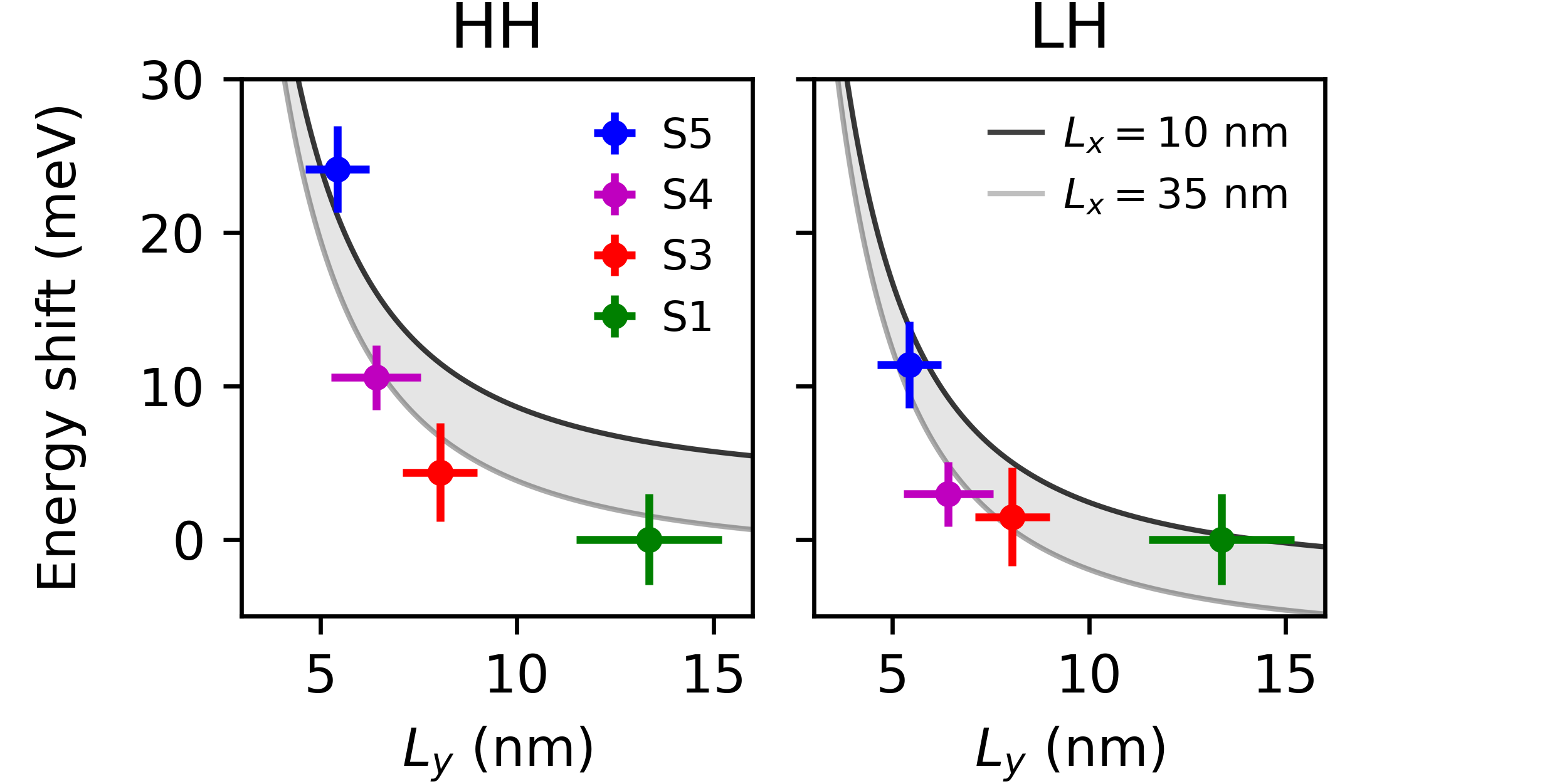}
\caption{Energy shift  for HH (left) and LH (right) exciton absorption peaks from Fig.\,\ref{fig1}(c) as a function of the small lateral size $L_y$. 
This  shift is considered with respect to  the HH and LH energies of sample S1. 
Shaded areas are  energy ranges obtained from Eq.\,(\ref{COM}) with $L_x$  between 10 and 35 nm.
}
\label{fig2}
\end{figure}

\begin{figure*}
\center
\includegraphics[width=1\textwidth]{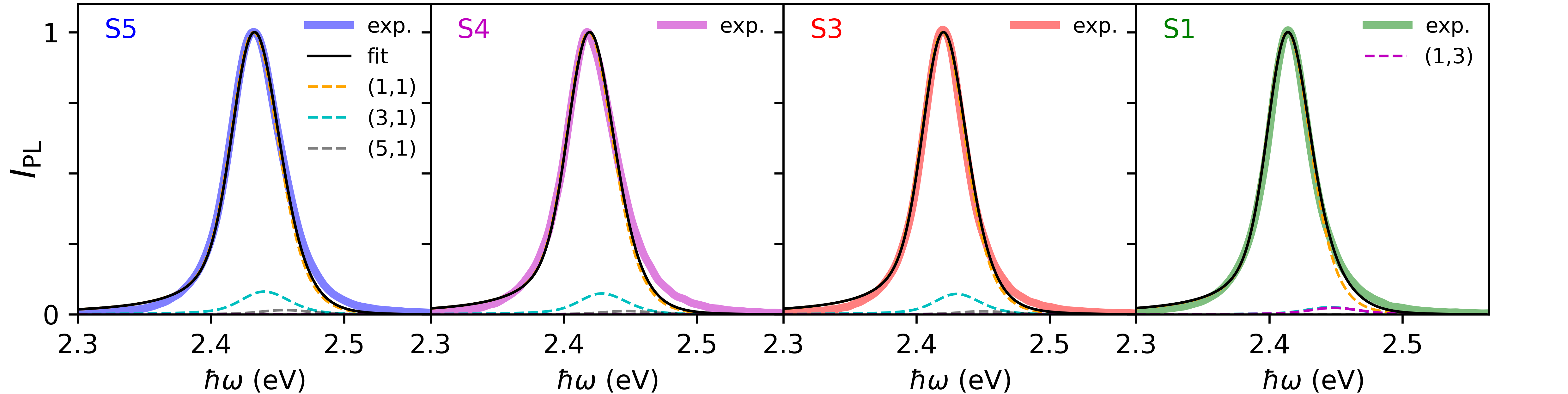}
\caption{Fits (black lines) of  experimental photoluminescence spectra (coloured thick lines) 
from 
our CdSe-NPLs with different lateral sizes (see Table\,\ref{tab}).
Our model considers the contribution of different  exciton COM motional  states (thin dashed coloured lines), with quantum numbers $\bm{n}=(n_x,n_y)$, whose energy and oscillator strength are  influenced by  the size distribution [Eq.\,(\ref{COM}) and (\ref{OS})]. 
 }
\label{fig3}
\end{figure*}

In order to interpret and explain  the observations listed above, we  now consider COM exciton states and their change with the  lateral dimensions of the NPLs. 
Since the NPL thickness $L_z$ is smaller than the bulk exciton Bohr radius, $a_{\textnormal B}^{\textnormal{3D}} = 5.4$\,nm \cite{IthurriaJACS:2011}, excitons experience strong quantum confinement along the $z$-direction \cite{Richter:2017,Rajadell:2017,Florez:2020}.
This strong  confinement increases the exciton binding energy and reduces the Bohr radius to  a quasi-2D value, $a_{\textnormal B}^{\textnormal{2D}} = 1.8\,$-\,$4.1$\,nm \cite{Tessier:2012,IthurriaJACS:2011,Achtstein:2012}, which is smaller than  the lateral sizes of  the NPLs investigated in this work. 
Hence, excitons  in a NPL are described by  approximately factorizing their perpendicular and in-plane wave functions as \cite{Zimmermann:1997,Richter:2017,Rajadell:2017}: $\Psi(z_e,z_h,\bm{r },\bm{R}) \, \simeq\,  u_{n_z}^e(z_e)u^h_{n_z}(z_h)\psi^{\textnormal{rel}}(\bm{r})\psi^{\textnormal{COM}}_{n_x,n_y}(\bm{R})$, where $z_{e,h}$ are the coordinates  of electrons ($e$) and holes ($h$) perpendicular to the plane of the NPL,  $\bm{r}= \bm{r}_e - \bm{r}_h$ and $\bm{R}= (m_e\bm{r}_e +m_h\bm{r}_h)/(m_e+m_h)$ are the relative and COM coordinates in the plane of the NPL, respectively,  with $m_e$ ($m_h$) being the electron (hole) effective mass, $m_e+m_h=M$ is the total exciton mass, while $n_i$ are quantum numbers along the $i$-th direction. 
The wave functions $u_{n_z}^{e,h}$ give rise to a confinement energy of both the electron and the hole of $E_z = E^{e}_{n_z=1}(L_z) + E^{h}_{n_z=1}(L_z)  = \pi^2\hbar^2/(2\mu L_z^2)$, where $\mu=m_em_h/(m_e+m_h)$ is the reduced effective mass of the $e$-$h$ pair.
The  wave function $\psi^{\text{rel}}(\bm{r})$ describes the relative or internal exciton wave function, which can be obtained by using the appropriate Coulomb potential. 
The COM wave function  $\psi^{\text{COM}}_{n_x,n_y}(\bm{R})$ is obtained by using the particle-in-a-box potential, resulting in COM exciton states with energy:
\begin{equation}
	E_{n_x,n_y}(L_x,L_y) = \frac{\pi^2\hbar^2}{2M}\Big[\Big(\frac{n_x}{L_x}\Big)^2 + \Big(\frac{n_y}{L_y}\Big)^2\Big].
	\label{COM}
\end{equation}
For convenience, in what follows we  use the notation $\bm{L}=(L_x,L_y)$ and $\bm{n}=(n_x,n_y)$, so that the dependence of the  energy in Eq.\,(\ref{COM}) on the lateral size and quantum number is denoted as $E_{\bm{n}}(\bm{L})$.

The validity of our approach is first verified by focusing on the energy shift of the HH and LH absorption peaks, as the small lateral size ($L_y$) is varied, as shown in Fig.\,\ref{fig2}. 
This shift is considered from the exciton peak energies (for HH and LH) of the larger NPLs (sample S1, green dots). 
Vertical and horizontal bars depict the indetermination of the experimental peak energies and $L_y$, respectively.  
The shaded area represents the range of calculated energies from Eq.\,(\ref{COM}) with $\bm{n}=(1,1)$, $m_e=0.27\,m_0$, the HH mass $m_{hh}=0.45\,m_0$,  the LH mass $m_{lh}=0.52\,m_0$ \cite{Benchamekh:2014}, and  $L_x$ between 10 and 35 nm.
The agreement between the calculated and experimental energy shift is a clear evidence of  the role of spatial lateral confinement on the exciton COM motion.\\

According to the particle-in-a-box model, the COM wave function for the in-plane motion of an exciton is given by: 
\begin{equation}
	\psi^{\textnormal{COM}}_{n_x,n_y}(x,y) = \sqrt{\frac{4}{L_x L_y}}\sin{\Big(\frac{n_x\pi x}{L_x}\Big)}\sin{\Big(\frac{n_y\pi y}{L_y}\Big)}.
	\label{PiB}
\end{equation}
The oscillator strength for photoexcitation from the electronic ground state is non-zero only for odd values of both $n_x$ and $n_y$, and is 
proportional 
to \cite{Zimmermann:1997}:
\begin{equation}
	f_{\bm{n}}(\bm{L}) = \Big| \int _0^{L_x} \hspace{-0.2cm} \int _0^{L_y} \hspace{-0.2cm} \text{d}x\text{d}y  \,\psi^{\textnormal{COM}}_{n_x,n_y}(x,y)\Big|^2 = \frac{4L_xL_y}{\pi^4 n_x^2n_y^2}.
\label{OS}
\end{equation}
The oscillator strength described by Eq.\,(\ref{OS}) is in line with the experimental observation of an enhanced  absorption cross section for larger NPLs \cite{Yeltik:2015}. 
Moreover, the inverse-square dependence on $\bm{n}$ agrees with the observed asymmetry at the high energy side of the HH exciton absorption peak, as shown in Fig.\,\ref{fig1}(c). \\

To model PL and absorption spectra of an ensemble of NPLs, it is essential to average over the known population distribution $D(\bm{L})$ of CdSe-NPLs in the sample, which is a Gaussian distribution reported in Fig. SM1 of the Supplemental Material. 
PL spectra are due to photon emission from HH excitons in COM motional states with quantum numbers $\bm{n}=(n_x,n_y )$. 
This leads to the following expression for the PL intensity at photon energy $\hbar\omega$: 
\begin{equation}
    \begin{split}
        {I}_{\textnormal{PL}}(\hbar \omega) & =   \sum_{\bm{n}}  \int \text{d} \bm{L} \, D(\bm{L})  f_{\bm{n}}(\bm{L})   e^{-\hbar\omega/k_{\textnormal{B}}T}    \\
                                              &    \cdot \mathcal{V}\left(\hbar \omega -E^{\textnormal{HH}}-E_{\bm{n}}^{\textnormal{HH}}(\bm{L}) ; \Gamma^{\textnormal{HH}}(\hbar\omega),\sigma^{\textnormal{HH}}\right).
    \end{split}
    \label{PL}
\end{equation}
The exponential factor is due to the classical Maxwell-Boltzmann (MB) distribution over thermalized COM motional states. 
The broadening of each of these states is modelled  by the Voigt distribution \cite{Achtstein:2016,Scott:2019}, $\mathcal{V}\left(\hbar \omega-E_0; \Gamma^{\textnormal{HH}}(\hbar\omega),\sigma^{\textnormal{HH}}\right)$, centered at $E_0=E^{\textnormal{HH}}+E_{\bm{n}}^{\textnormal{HH}}(\bm{L})$, where $E^{\textnormal{HH}}$ is the exciton state with zero COM energy, which is determined also by the $z$-confinement  and Stokes shift.
The term  $E_{\bm{n}}^{\textnormal{HH}}(\bm{L})$ is calculated from Eq.\,(\ref{COM}).
$\Gamma^{\textnormal{HH}}$ and $\sigma^{\textnormal{HH}}$ are the linewidths of the Lorentzian  and Gaussian distribution defining $\mathcal{V}$, respectively.
The former accounts for the exciton-phonon interaction which, according to the Urbach rule \cite{HaugAndKoch},  is given by: 
$\Gamma^{\textnormal{HH}}(\hbar\omega) = \Gamma_0^{\textnormal{HH}}/ (e^{[E^{\textnormal{HH}}+E_{\bm{n}}^{\textnormal{HH}}(\bm{L})-\hbar\omega]/k_BT}  +1)$. 
$\sigma^{\textnormal{HH}}$ is  here attributed  to the inhomogeneous broadening due to structural, size-independent disorder; e.g. non-flatness of NPLs, effects of surface ligands on the NPLs, or to the deviation of the shape of the NPLs from perfect rectangles.
The only fit parameters for modeling  PL spectra from Eq.\,(\ref{PL}) for all samples are $E^{\textnormal{HH}}$, $\Gamma^{\textnormal{HH}}_0$, and $\sigma^{\textnormal{HH}}$.

\begin{figure}
\includegraphics[width=0.5\textwidth]{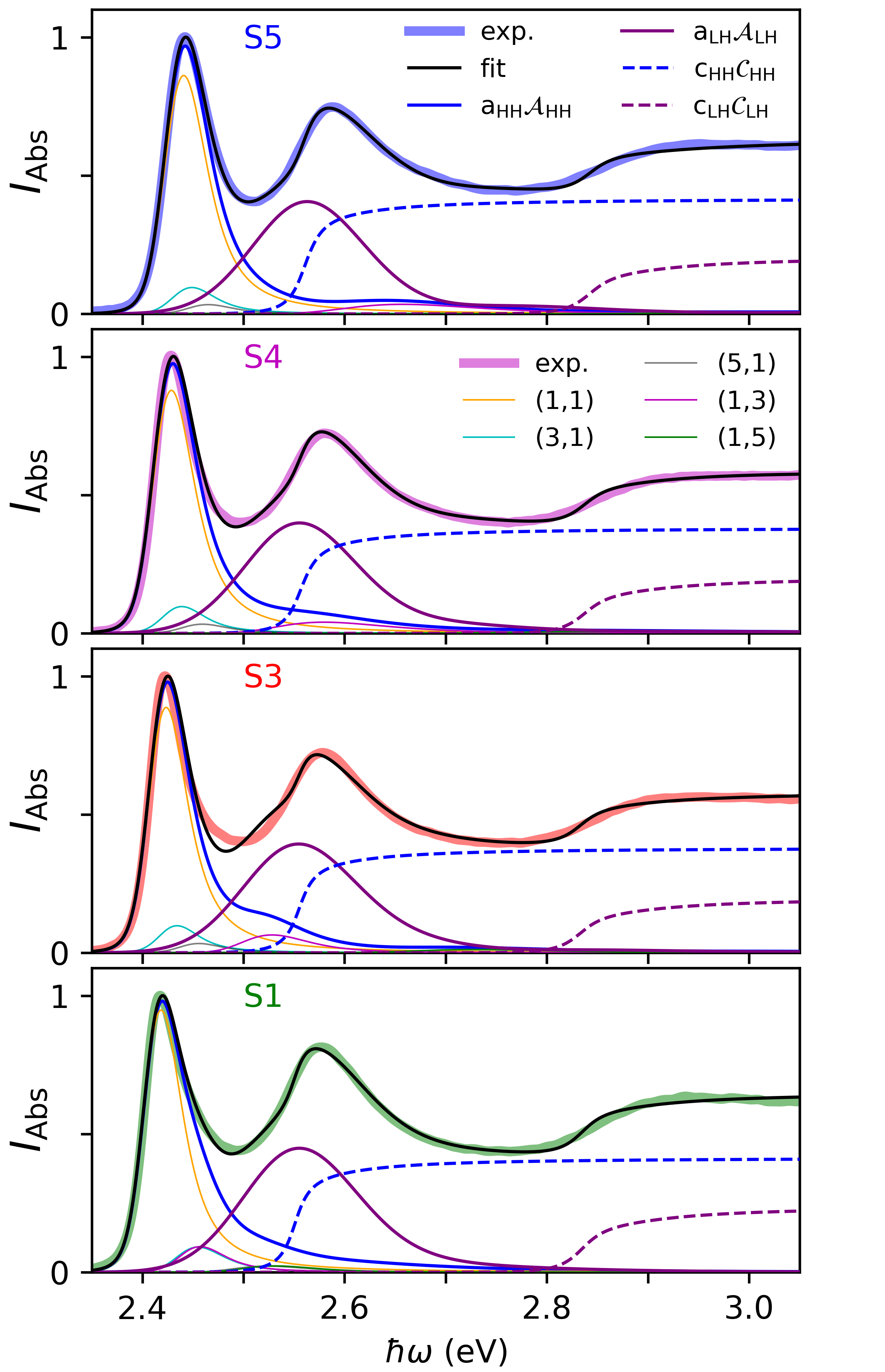}
\caption{Fits (black lines)  of our model given by  Eqs.\,(\ref{ExAbs}), (\ref{ContAbs}) and (\ref{totAbs}) to the experimental absorption spectra. 
The presence of   exciton COM motional  states,  $\bm{n}=(n_x,n_y)$ (depicted by coloured thin lines), and their dependence on the lateral sizes of the NPLs, explains and reproduces the change in the asymmetry at the high-energy side of the HH absorption peak.
 }\label{fig4}
\end{figure}

Fig.\,\ref{fig3} shows  the obtained PL fits  (black thin lines) which  agree  very well  with the experimental  spectra (thick colored lines) for every sample. 
The fits are obtained with $\Gamma^{\textnormal{HH}}_0 = 26\pm 1$\,meV and $\sigma^{\textnormal{HH}}=14\pm 1$\,meV for all samples.
The contribution of different states is also depicted (thin lines). 
For rectangular NPLs ($r_l>1$),  besides the main PL from (1,1) states, the contribution at higher energies is due to (3,1)  (magenta)  and (5,1) (grey) states.
The small contribution from the (5,1) states is due to the decrease of the oscillator strength with $\bm{n}$, see  Eq.\,(\ref{OS}), as well as to the MB distribution.
Due to the square shape of the NPLs in the S1 sample  ($r_l=1$), the (3,1) and (1,3) (cyan)  states contribute with a similar oscillator strength and energy. 
Since $\Gamma^{\textnormal{HH}}_0$ and $\sigma^{\textnormal{HH}}$ are the same for all samples, the decreasing experimental PL linewidth, by going from sample S5 to sample S1, is explained by a reduction of the effects of the lateral size distribution of the NPLs via   $D(\bm{L})$ and, in turn $f_{\bm{n}}(\bm{L})$.\\

The exciton absorption is modeled similarly to the PL and is given by:
\begin{equation}
    \begin{split}
        \mathcal{A}^J(\hbar \omega) & = \sum_{\bm{n}} \int \text{d} \bm{L} \, D(\bm{L})  f_{\bm{n}}(\bm{L})  \\
                                                      & \cdot \mathcal{V}\left(\hbar \omega  -  E^{J} - E_{\bm{n}}^{{J}}(\bm{L}); \Gamma^{{J}}(\hbar\omega),\sigma^{{J}}\right),
    \end{split}
    \label{ExAbs}
\end{equation}
where $J=$ HH, LH.
Note, that the MB distribution does not appear in  Eq.\,(\ref{ExAbs}), since the initial state for absorption is the electronic ground state of a NPL. 
Transitions  from the valence to the conduction band (continuum) need also to be taken into account and are calculated as:
\begin{equation}
    \begin{split}
        \mathcal{C}^J(\hbar \omega) = \sum_{\bm{n}} \int &\text{d} \bm{L} \, D(\bm{L}) f_{\bm{n}}(\bm{L}) \\
        & \cdot \int_{-\infty}^\infty \hspace{-0.3cm} \text{d} \epsilon_0 \,  \mathcal{V}\left(\hbar \omega -\epsilon_0 ;  \Gamma^{J}_c(\hbar\omega),\sigma^{J}_c\right)  \\
                                                      &\cdot \Theta\left(\epsilon_0 - {E}_{X}^J- E^J - E_{\bm{n}}^J(\bm{L}) \right), 
    \end{split}
       \label{ContAbs}
\end{equation}
where $\Gamma^{{J}}_c$ and $\sigma^{J}_c$ are the Lorentzian and Gaussian linewidths of the  continuum, respectively, $\Theta(x)$ is the step-function (that is, $\Theta(x) = 1$ for $x>0$ and zero otherwise), and ${E}_{X}^J$ is the exciton binding energy, which is defined to be positive. 
By means of Eqs.\,(\ref{ExAbs}) and (\ref{ContAbs}), the absorption due to all above-mentioned states is:
\begin{equation}
    \begin{split}
 {I}_{\textnormal{Abs}}(\hbar \omega)  & = a^{\textnormal{HH}} \mathcal{A}^{\textnormal{HH}}(\hbar \omega) 
                                 + a^{\textnormal{LH}}   \mathcal{A}^{\textnormal{LH}}(\hbar \omega) \\
                             & + c^{\textnormal{HH}}   \mathcal{C}^{\textnormal{HH}}(\hbar \omega) + c^{\textnormal{LH}}   \mathcal{C}^{\textnormal{LH}}(\hbar \omega) ,
                             \label{totAbs}
    \end{split}
\end{equation}
where  each  contribution is weighted with  fitted coefficients $a^J$ and $c^J$ that take into account the optical dipole moments involving the relative motion of electrons and holes in the exciton and continuum states, respectively.  
The values of $E^{\textnormal{HH}}$, $\Gamma^{\textnormal{HH}}_0$, and $\sigma^{\textnormal{HH}}$ obtained by fits to  the PL spectra are also used to calculate the contribution from $ \mathcal{A}^{\textnormal{HH}}(\hbar \omega)$. 
For LH excitons,  $E^{\textnormal{LH}}$, $\Gamma^{\textnormal{LH}}_0$, and $\sigma^{\textnormal{LH}}$ are obtained  from fits to the absorption spectra.
The parameters  $\Gamma^{{J}}_c$,  and ${E}_{X}^J$  are also fitted, while $\sigma^J_c$ is taken to be zero for simplicity, as we expect delocalized continuum states to be less affected by local disorder. 

Fig.\,\ref{fig4} shows the obtained fits (black thin lines)  and  the experimental absorption (colored thick lines). 
The total contribution from HH (blue solid lines) and LH excitons (magenta solid lines)  are also shown together with the corresponding continuum contributions (dashed lines).
Thin  colored  lines depict different $\bm{n}$-states for HH excitons. 
Our model thus reproduces very well the entire absorption spectra for different lateral sizes of NPLs.
The fitting procedure results in exciton binding energies equal to $E_{X}^{\textnormal{HH}} = 135 \pm2$\,meV and $E_{X}^{\textnormal{LH}} = 301 \pm4$\,meV, which are independent of the lateral sizes to within the uncertainty. 
This is to be expected, since  $L_x,L_y>a_B^{2D}$ for all samples, and consistent with our separable ansatz for the exciton wave functions on which our model is based. 
Other  fitting parameters are listed in Table\,SMI of  the Supplemental Material.  
Note that all fit parameters do not depend on the lateral sizes of the NPLs, as expected. 
The asymmetry of the HH peak clearly depends on the presence of different exciton COM motional states.   
The (1,3) state redshifts on going from sample S5 to sample S1 due to the increase of the lateral size $L_y$.
In addition, the enhanced HH asymmetry for the S1 sample results from the similar contribution of (1,3) and (3,1) states.\\

In conclusion, we demonstrated that effects of the lateral size of CdSe NPLs on PL and optical absorption spectra can be fully attributed to the quantized 
exciton COM motion.
The experimental spectra can be reproduced theoretically on the basis of the simple particle-in-a-box model for the translational motion of the exciton within a NPL. 
Importantly, it was not necessary to take into account size-dependent effects of phonons on PL and absorption spectra. 
Our results show that variation of the lateral size of (quasi) two-dimensional materials provides a tool to tune their excitonic and therefore optoelectronic properties in addition to modifying the thickness. 
The new insights also open the road to studying nontrivial quantum effects such as the Berry phase on the COM motion of topological excitons in NPLs consisting of topological insulator materials. \\

\begin{acknowledgments}
L.\,D.\,A.\,S.\,and M.\,F.\,thank A.\,W.\,Achtstein for the very helpful discussions.
This work is part of the research program TOP-grants with project number 715.016.002, which is financed by the Netherlands Organization for Scientific Research (NWO). 
\end{acknowledgments}

\end{document}